\documentclass{mn2e}

\title[Erratum: Cooling flows, black holes and the luminosities and colours of galaxies]
{Erratum - The many lives of AGN: cooling flows, black holes and the luminosities
and colours of galaxies}

\author[Croton et al.]{
\parbox[t]{\textwidth}{
Darren J. Croton$^{1}$,
Volker Springel$^{1}$.
Simon D. M. White$^{1}$,
G. De Lucia$^1$,
C. S. Frenk$^2$,
L. Gao$^1$,
A. Jenkins$^2$,
G. Kauffmann$^1$,
J. F. Navarro$^3$,
N. Yoshida$^4$
}
\vspace*{6pt} \\ 
$^1$Max-Planck-Institut f\"ur Astrophysik, D-85740 Garching, Germany.\\
$^2$Institute for Computational Cosmology, Physics Department, Durham. U.K.\\
$^3$Department of Physics and Astronomy, University of Victoria, B.C., Canada.\\
$^4$Department of Physics, Nagoya University, Chikusa-ku, Nagoya 464-8602, Japan\\
\vspace{-0.5cm} 
}

\date{Accepted ---. Received ---;in original form ---}
\pubyear{2005}

\begin{document}

\maketitle


In preparing the script to plot the Tully-Fisher relation in Figure~6
of our paper ``The many lives of active galactic nuclei: cooling
flows, black holes and the luminosities and colours of galaxies''
(Croton et al. 2006, MNRAS, 365, 11) we inadvertently loaded the
incorrect column from our database with the result that the proxy
$V_{\rm c}$ is not the virial velocity of the parent dark matter halo,
$V_{\rm vir}$, as labeled, but is instead the maximum circular
velocity of the dark matter halo, $V_{\rm max}$. The maximum halo
circular velocity is a much better estimate of the disk $V_{\rm c}$
than is $V_{\rm vir}$.  This confusion influenced the discussion of
the Tully-Fisher relation in our paper. In fact, Figure~6 demonstrates
that it is possible to simultaneously reproduce both the local
Tully-Fisher relation and luminosity function using semi-analytic
techniques applied to the standard $\Lambda$CDM cosmology, thus
contradicting previous studies of this issue and our own discussion in
Section~3.6. We will explore this promising aspect of our galaxy
formation model in a future paper.

\section*{Acknowledgments}
\label{acknowledgements}

The authors would like to thank Qi Guo for pointing out this error in
our manuscript.

\label{lastpage}

\end{document}